# ACTIVE AND COOPERATIVE LEARNING PATHS IN THE PIGELLETO'S SUMMER SCHOOL OF PHYSICS

**Roberto Benedetti, Emilio Mariotti, Vera Montalbano,** *Phys. Department, University of Siena*
**Antonella Porri,** *Regional Scholastic Office of Tuscany - Arezzo territorial area*

**Abstract**
Since 2006, the Pigelleto's Summer School of Physics is an important appointment for orienting students toward physics. It is organized as a full immersion school on actual topics in physics or in fields rarely pursued in high school, i.e. quantum mechanics, new materials, energy resources. The students, usually forty, are engaged in many activities in laboratory and forced to become active participants.
Furthermore, they are encouraged in cooperating in small groups in order to present and share the achieved results. In the last years, the school became a training opportunity for younger teachers which are involved in programming and realization of selected activities. The laboratory activities with students are usually supervised by a young and an expert teacher in order to fix the correct methodology.

## 1. The Pigelleto's Summer School of Physics

In the last years, many Italian Universities are involved in a large national project[1] in order to enhance the interest of high school students towards scientific degrees, in particular in Physics, Mathematics and Chemistry.

In this context, we started to organize a full immersion summer school of physics in the Pigelleto's Natural Reserve, on the south east side of Mount Amiata in the province of Siena. Our purpose is offering to really motivated students an opportunity of testing the scientific method, the laboratory experience in a stimulating context, by deepening an interesting and relevant topic in order to orienting them towards physics.

Forty students from high school are selected by their teachers in a wide network of schools within the National Plan for Science Degree[2] and come from the south Tuscany (Arezzo, Siena. Grosseto).

The school is usually held in the beginning of September and lasts for four days. The 2011 edition was titled *Thousand and one energy: from sun to Fukushima* (Di Renzone 2012). Some previous editions were *Light, colour, sky: how and why we see the world* (Porri 2008a,b), *Store, convert, save, transfer, measure energy, and more…*(Porri 2008a,b), *The achievements of modern physics* (Porri 2010, Benedetti 2011), *Exploring the physics of materials* (Di Renzone 2011, Montalbano 2011).

Topics are chosen so that students are involved in activities rarely pursued in high school, aspects and relationship with society are underlined and discussed .

In the morning, we usually propose lectures in which, by stimulating the active involvement, we give the necessary background for the following activities in laboratory. In the afternoon, small groups of students from different school and classes are engaged in laboratories activities, playing an active role. Every group is supported by one or two teachers that are available for discussions and hints. We propose different laboratories, usually eight in different places. Sometimes they focus on the same physics phenomena explored by different point of view. Every group has the task of preparing a brief presentation in which describing to other students what it has been learned in lab. In the evening, a sky observation is proposed. If it is cloudy, we can propose an indoor problem solving.

In the last years, the school became a training opportunity for younger teachers which are involved in programming and realization of selected activities.

Many educational choices are made in programming the summer school:
- main target and methodologies are discussed and selected with the teachers involved in PLS

---

[1] Progetto Lauree Scientifiche, i. e. Scientific Degree Project from 2006 to 2009
[2] i.e. Piano nazionale Lauree Scientifiche, which continues the legacy of the previous project since 2010

- almost all laboratories are made with poor materials or educational devices provided by some schools
- the groups are inhomogeneous and formed by following the teachers suggestions in order to promote the best collaboration
- the main topic is related to all activities and must be not trivial
- almost all laboratories lead to at least one measure and its error evaluation.

Active learning (Bonwell 1991, Paulson 2011) plays a central role throughout all school activities. During the morning lectures, students are engaged in more activities than just listening. They are involved in discussion of or thinking about issue before any theory is presented in lecture or after several conflicting theories have been presented. We often present them a paradox or a puzzle involving the concept at issue and have them struggle towards a solution, by forcing the students *to work it out* without some authority's solution.

In the laboratory activities, they are encouraged in cooperating in small groups in order to present and share the achieved results. The laboratories are organized in such a way that cooperative learning is implemented. A closer analysis of our laboratories shows that all requests for achieving a cooperative learning are satisfied (Curseo 1992, Johnson 1999), such as positive interdependence, individual accountability, face-to-face promotive interaction, social skills and group processing.

## 2. Some learning path

Two examples of learning paths made in Pigelleto's summer school are shown in Tab.1.

We emphasize that it was always planned in the lab at least an activity that we can call *How it works*. This type of activity is very stimulating for students that literally *discover* the underlying physics, as expected in a context-based approach (Kortland 2005).

*Table 1: Two examples of learning paths*

| The achievements of modern physics<br>$7^{th}$ - $10^{th}$ September 2009 | Thousand and one energy:<br>from sun to Fukushima<br>$5^{th}$ - $8^{th}$ September 2011 |
|---|---|
| *Morning lectures* | *Morning lectures* |
| On the validity of a physical theory<br>On concepts of classical physics disproved by modern physics<br>Atomic models: from classical to modern ones<br>How laser works<br>Radioactivity, nuclei and surrounding<br>On nuclear energy and energy resources | On Energy<br>Energy from nuclei<br>Energy from sky and sun<br>Energy from the ground<br>On photovoltaic panels<br>On fuel cells<br>Conductors and superconductors |
| *Laboratories* | *Laboratories* [3] |
| How it works: a spectroscope<br>Measurement of Plank's constant with a LED[4]<br>Photoelectric effect[5]<br>We characterize a diode laser<br>We characterize the natural radioactivity<br>Measure of the speed of rotation of a star[6] | How it works: a photovoltaic panel<br>How it works: Stirling machine, solar oven, coffeepot<br>Measurement of the mechanical equivalent of heat<br>Electromagnetic induction and energy dissipation<br>Newton's cradles and Yo-Yos<br>Radioactivity background vs. small Uranium sources |

---

[3] (Di Renzone 2012)
[4] (Benedetti 2011)
[5] (Benedetti 2011)
[6] (Porri 2010)

## 3. Communicate Physics

The oral talk in which students share what they have learned in laboratory is stressed as the central part of the school. Students are not used to speak in public and fear this moment but many of them respond well to the challenge. They can choose the form and the means utilized in the communication. Everyone can talk or a group can delegate a speaker, they can present the data in the blackboard or with slides. Furthermore, they can show devices or material for better explain the topic. At the end a brief discussion is performed between the communicators and the public.

In the beginning the communications were proposed in order to share the results of each lab to everyone. It was a necessity because we haven't enough time and space to perform all experiments with everyone. In the years, these communications are became the central activity that stimulates students to be active and cooperative in the learning path.

## 4. Conclusions

The Pigelleto's Summer School of Physics plays a central role for orienting students toward physics in our territorial area. Our main goal is the excellent feedback that returns from students, their families and teachers. In the last years, the applications are increased up to double the available positions. In our opinion, the main reason of this success is the active and cooperative learning of interesting topics in physics that students have experienced during the summer school.